\documentclass[prl,twocolumn,aps,superscriptaddress,preprintnumbers]{revtex4}
\usepackage{graphicx,xcolor}
\usepackage{amsfonts}
\usepackage{amssymb}
\usepackage{amsmath}
\usepackage{dsfont}
\usepackage{hyperref}
\usepackage{slashed}
\usepackage{empheq}
\usepackage{multirow}
\usepackage{fancyhdr}
\usepackage{appendix}
\usepackage{subfigure}
\usepackage{times}

\DeclareMathOperator{\hz}{h\relax{\kern-.15em}z}
\DeclareMathOperator{\pz}{\psi\relax{\kern-.15em}z}

\usepackage[english]{babel}
\usepackage{amssymb}
\usepackage{amsfonts}
\usepackage{amsmath}
\usepackage{graphicx}
\usepackage{epsfig}
\usepackage{psfrag}
\newcommand{\be}{\begin{equation}} \newcommand{\ee}{\end{equation}}
\newcommand{\bea}{\begin{eqnarray}} \newcommand{\eea}{\end{eqnarray}}
\newcommand{\beann}{\begin{eqnarray*}}  \newcommand{\eeann}{\end{eqnarray*}}
\newcommand{\bfig}{\begin{figure}} \newcommand{\efig}{\end{figure}}
\newcommand{\ba}{\begin{array}} \newcommand{\ea}{\end{array}}
\newcommand{\bcen}{\begin{center}} \newcommand{\ecen}{\end{center}}
\newcommand{\btab}{\begin{tabular}} \newcommand{\etab}{\end{tabular}}

\newcommand{\matt}{\left ( \begin{array}{ccc}}
    \newcommand{\ematt}{\end{array} \right )} \newcommand{\matf}{\left ( \begin{array}{cccc}}
    \newcommand{\ematf}{\end{array} \right )} \newcommand{\vect}{\left ( \begin{array}{c}}
    \newcommand{\evect}{\end{array} \right )}    \def\beqn{\begin{eqnarray}}
 \def\eeqn{\end{eqnarray}}  
     
\newtheorem{Proposition}{Proposition}[section]

\newtheorem{Theorem}{Theorem}[section]
\newtheorem{Lemma}{Lemma}[section]
\newtheorem{Corrolary}{Corrolary}[section]

\newcommand{\bp}{\begin{Proposition}}	\newcommand{\ep}{\end{Proposition}}
\newcommand{\bt}{\begin{Theorem}}	\newcommand{\et}{\end{Theorem}}
\newcommand{\bl}{\begin{Lemma}}		\newcommand{\el}{\end{Lemma}}
\newcommand{\bc}{\begin{Corrolary}}	\newcommand{\ec}{\end{Corrolary}}

\begin{document}

\preprint{MPP-2015-151}
\preprint{MCTP-15-08}

\title{The Holographic Disorder-Driven  Superconductor-Metal Transition}

\author{D. Are\'an}\email{darean@mpp.mpg.de}
\affiliation{Max-Planck-Institut f\"ur Physik (Werner-Heisenberg-Institut,  F\"ohringer Ring 6, D-80805, Munich, Germany }

\author{L. A. Pando Zayas}\email{lpandoz@umich.edu}
\affiliation{The Abdus Salam International Centre for Theoretical Physics, Strada Costiera 11, I 34014 Trieste, Italy}
\affiliation{Michigan Center for Theoretical Physics, Randall Laboratory of Physics, University of Michigan, Ann Arbor, MI 48109, USA}

\author{I. Salazar Landea}\email{peznacho@gmail.com}
\affiliation{ Instituto de F\'\i sica La Plata  and Departamento de F\'\i sica Universidad Nacional de La Plata, CC 67,
1900 La Plata, Argentina}

\author{A. Scardicchio}\email{ascardic@ictp.it}
\affiliation{The Abdus Salam International Centre for Theoretical Physics, Strada Costiera 11, I 34014 Trieste, Italy}
\affiliation{INFN, Sezione di Trieste, Via Valerio 2, I 34127, Trieste, Italy}

\begin{abstract}
We implement the effects of disorder on a holographic superconductor by introducing
a random chemical potential on the boundary. We demonstrate explicitly that increasing disorder leads to the formation of islands where the superconducting order is enhanced and subsequently to the transition to a metal.
We study the behavior of the superfluid density and of the conductivity as a function of the strength of disorder. We find explanations for various marked features in the conductivities in terms of hydrodynamic quasinormal modes  of the holographic superconductors. These identifications plus a particular disorder-dependent  spectral weight shift in the conductivity point to a signature of the Higgs mode in the context of disordered holographic superconductors. 
We observe that the behavior of the order parameter close to the transition is not mean-field type as in the clean case, rather we find robust agreement with $\exp(- A\, |T-T_c|^{-\nu})$, with $\nu =1.03\pm 0.02 $ for this disorder-driven smeared transition.
\end{abstract}

\pacs{ }

\maketitle


{\bf  Introduction: } The suppression of conductivity due to disorder, known as Anderson localization,  is one of the most striking transport phenomena that involves quantum behavior \cite{Anderson:1958vr}. The application of  localization ideas to superconductors was for a long time governed by Anderson's theorem stating that  nonmagnetic impurities have no
significant effect on the superconducting transition since Cooper pairs are formed from time reversed
eigenstates, which included disorder \cite{Anderson195926}. Anderson's
idea applies only to weakly disordered systems (with extended electronic states) and weak interactions; the role of interactions has been further discussed by Ma and Lee in \cite{PhysRevB.32.5658}. The mechanism put forward in \cite{PhysRevB.32.5658} states that strong disorder gives rise to spatial fluctuations of the order parameter along with its suppression in comparison to its value in the clean system. The existence of spatial fluctuations of the order parameter has recently been corroborated and argued to be the central mechanism in the 
superconductor-insulator transition \cite{PhysRevLett.87.087001,PhysRevLett.81.3940}. 
Various experiments have indicated that high-$T_c$ superconductors are intrinsically 
disordered~\cite{Pan_Nature,Renner_Nature} emphasizing the need for a better understanding of the 
interplay between superconductivity, strong interactions and disorder. In this manuscript we tackle
this interplay using holographic methods.

The Anti de Sitter/Conformal Field Theory (AdS/CFT) correspondence has already succeeded in constructing
holographic versions of superconductors
\cite{Hartnoll:2008vx,Hartnoll:2008kx} (for reviews see \cite{Hartnoll:2009sz,Horowitz:2010gk}).
Some holographic models have been argued to be relevant to describe materials like the cuprates
\cite{Hartnoll:2009ns,Sachdev:2011wg} since
their dynamics is believed to be largely governed by their proximity to a quantum critical point
\cite{sachdev2011}.

In previous work \cite{Arean:2013mta,Arean:2014oaa} we investigated mild disorder in holographic 
superconductors  and found an enhancement of the superconducting order parameter. 
Other recent works studying disorder holographically include \cite{Zeng:2013yoa,Hartnoll:2014cua,O'Keeffe:2015awa}. 
Here we explore the regime of large disorder and provide a detailed description of the superconductor-metal 
transition. Although we work in a setup where the normal phase describes a metal, it 
is also possible to consider insulating setups through AdS/CFT \cite{Nishioka:2009zj,Donos:2012js,Donos:2014oha}. 

In this paper we report various properties of the superconducting transition 
including the averaged and spatially resolved behavior of the condensate, superfluid density 
and conductivity and provide an explanation in terms of the relevant hydrodynamic modes. 
We observe that the disorder-driven transition is smeared and the averaged order parameter 
(here the average $\langle\cdot\rangle$ is over the spatial coordinate or, equivalently, disorder realizations) 
vanishes very quickly as $\langle {\cal O}\rangle \sim \exp\left(- A|T-T_c|^{-\nu}\right)$; 
we find $\nu = 1.03\pm 0.02$. We expect that our explicit result for this exponent will stimulate 
alternative approaches to disorder-driven transitions to provide a quantitative characterization of the transition.


{\bf   Disordered holographic superconductor:}
To build a noisy holographic $s$-wave superconductor in 2+1
 dimensions we  consider, following  \cite{Hartnoll:2008vx}, the dynamics of a Maxwell field and a charged scalar in a fixed metric background:
\bea S=\int d^4
x\,\sqrt{-g}\left(-{1\over4}F_{ab}\,F^{ab}-(D_\mu\Psi)(D^\mu\Psi)^\dagger-m^2\Psi^\dagger\Psi
\right). \nonumber  \eea
The system is studied on the  Schwarzschild-$AdS_4$
metric: \bea ds^2&=&{1\over z^2}\left(-f(z)dt^2+{dz^2\over
f(z)}+dx^2+dy^2
\right),\nonumber \\
f(z)&=&1-z^3\,,
\eea
where we have set the radius of AdS, $R=1$, and the horizon at $z_h=1$.
This system is dual to a 2+1 CFT living on the boundary of $AdS_4$,
and the $U(1)$ gauge field realizes a conserved current. The temperature of the black hole is identified
with that of the field theory, and by fixing the horizon radius we are making use of the rescaling symmetry
of our theory to work in units of temperature.
We take the following (consistent) Ansatz for the matter fields:
\bea
&&\Psi(x,z)=\psi(x,z)\,,\quad  A=\phi(x,z)\,dt,
\eea
where $\psi(x,z)\in {\mathbb R}$.
The resulting equations of motion read
{\small
\bea
 \hspace*{-0.5cm} &&\partial_z^2\phi+\frac{1}{f}\,\partial_x^2\phi-{2\psi^2\over z^2\,f}\,\phi=0\,,\label{eomphi}\\
 \hspace*{-0.5cm} &&\partial_z^2\psi+\frac{1}{f}\,\partial_x^2\psi+\left({f'\over f}-{2\over z}\right)\partial_z\psi+\frac{1}{f^2}\left(\phi^2-{m^2\,f\over z^2}
\right)\psi=0\,. \nonumber\\ \label{eompsi}
\eea}
In what follows we choose the scalar mass  $m^2 = -2$, corresponding to a dual operator of conformal dimension $\Delta=2$.

The main rule for reading the  AdS/CFT dictionary states that  field theory information is extracted from the 
boundary values of the gravity fields. 
The UV ($z=0$) asymptotics of Eqs.~(\ref{eomphi},\ref{eompsi}) lead to
\bea
&&\phi(x,z)=\mu(x)+\rho(x)\,z+\phi^{(2)}(x)\,z^2+o(z^3)\,, \label{Eq:phi}\\
&&\psi(x,z)=\psi^{(1)}(x)\,z+\psi^{(2)}(x)\,z^2+o(z^3)\,, \label{Eq:psi}
\eea
where $\mu(x)$ and $\rho(x)$ correspond, in the dual field theory, to space-dependent chemical potential and 
charge density respectively.  The functions $\psi^{(1)}(x)$
and $\psi^{(2)}(x)$ are identified, under the duality, with the source and VEV of an operator of dimension 2. 
Imposing $\psi^{(1)}(x)=0$ in Eq.~(\ref{Eq:psi})  corresponds to spontaneous breaking of the $U(1)$ symmetry 
with order parameter ${\cal O}\propto\psi^{(2)}(x)$. 
In the IR $(z\sim z_h=1)$ regularity implies that $A_t$ vanishes at the horizon.


To mimic the choice of random on-site potential used originally by Anderson 
in \cite{Anderson:1958vr} we implement disorder by introducing a noisy chemical potential:
\bea
\mu(x)&=&\mu_0+{\mu_0\over25}\,w\,\sum_{k=k_0}^{k_*}\,\cos(k\,x+\delta_k)\,,
\label{noisefunc}
\eea
where $\delta_k$ is a random phase for each $k$, and $w$ is a free parameter that determines the strength
of the disorder.  
Notice that the system is homogeneous along the remaining spacelike direction $y$.
We discretize the space, and impose periodic
boundary conditions in the $x$ direction, leading to $k$ with values:
\be
k_n={2\pi\over L}\,(n+1)\quad {\rm with}\quad 0\leq n< N=\frac{k_*}{k_0}\,,
\ee
where $L$ is the length in the $x$ direction of our cylindrical space. 
Our noise is  a truncated version of Gaussian white noise
where the highest wave number $k_*$ takes the role of the inverse of the correlation length for the
chemical potential. More details on the properties of this choice of disorder can be found in \cite{Arean:2014oaa}.

Summarizing, we think of $k_0$ as the inverse system size and of $k_*$  as the inverse correlation length,
and we work in the regime $k_0/T \ll 1$.
More precisely, for most of the  simulations we take $L=20\pi$, and $k_*=1$, but we checked the stability of 
the results for lengths up to $L=80\pi$. 

To find solutions describing disordered superconducting states we integrate 
the equations of motion (\ref{eomphi}, \ref{eompsi}) for different realizations of disorder characterized 
by sets of the random phases $\delta_k$. We define the expectation values for the different observables by
averaging over different realizations. Typical plots correspond to about fifty realizations.


{\bf The order parameter and superconducting islands:} 
Let us focus on the behavior of the order parameter as a function of the chemical potential $\mu$ 
and the dimensionless strength of disorder $w$. We will pay special attention to the minimum of the 
condensate ${\cal O}_{\rm min}$ throughout the sample which we expect to be related with the DC conductivity.

\begin{figure}[htb]
\centering
\includegraphics[width=0.40\textwidth]{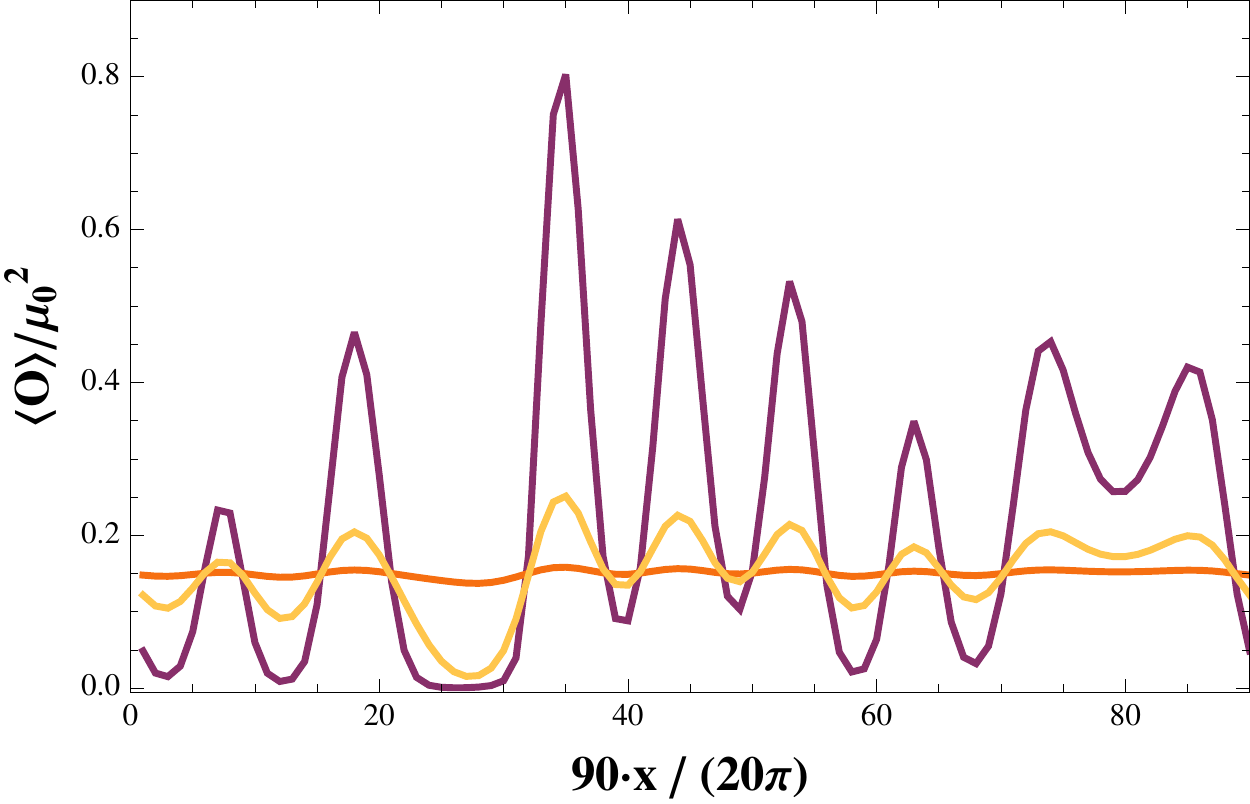}
\caption{The condensate as a function of the coordinate $x$ for three realizations with strength  $w=0.1, 0.9, 3,$
and the same set ${\delta_k}$ (orange, yellow and purple lines respectively), at a temperature $T=0.81T_c^{w=0}$.
This plot shows the appearance of islands, that is, of spatial fluctuations in the condensate. }
   \label{fig:Islands}
 \end{figure}

In \cite{Arean:2013mta,Arean:2014oaa} 
it was shown that noise enhances the spatial average of the order parameter. However,
already Ma and Lee found in \cite{PhysRevB.32.5658} that strong disorder gives rise to spatial 
fluctuations of the order parameter along with its suppression in comparison with its value in the 
clean/homogeneous system.
In Fig.~\ref{fig:Islands} we plot the spatial dependence of the order parameter 
at a temperature $T=0.81T_c^{w=0}$, where $T_c^{w=0}$ is the homogeneous critical temperature.
We find that for strong enough noise regions
akin to islands appear in the system. 

The enhancement of superconducting properties reported in \cite{Arean:2013mta,Arean:2014oaa} has some precedent
in the 
condensed matter literature. The role of superconducting islands was already explicitly mentioned in  
\cite{PhysRevLett.74.2800}, and further analyzed in \cite{PhysRevLett.87.087001}.
In that work it was shown that  for each realization of disorder, there are spatial regions where the local
upper critical field 
exceeds the system-wide average value. These regions form superconducting islands weakly coupled via the
Josephson 
effect. At low temperatures, proximity coupling is long ranged and, thus,  global superconductivity may
be established in the 
system.  Similar arguments were also advanced in \cite{PhysRevLett.81.3940}.
We see that precisely this mechanism seems to be at play in our holographic model.

\begin{figure}[htb]
   \centering
   \includegraphics[width=0.40\textwidth]{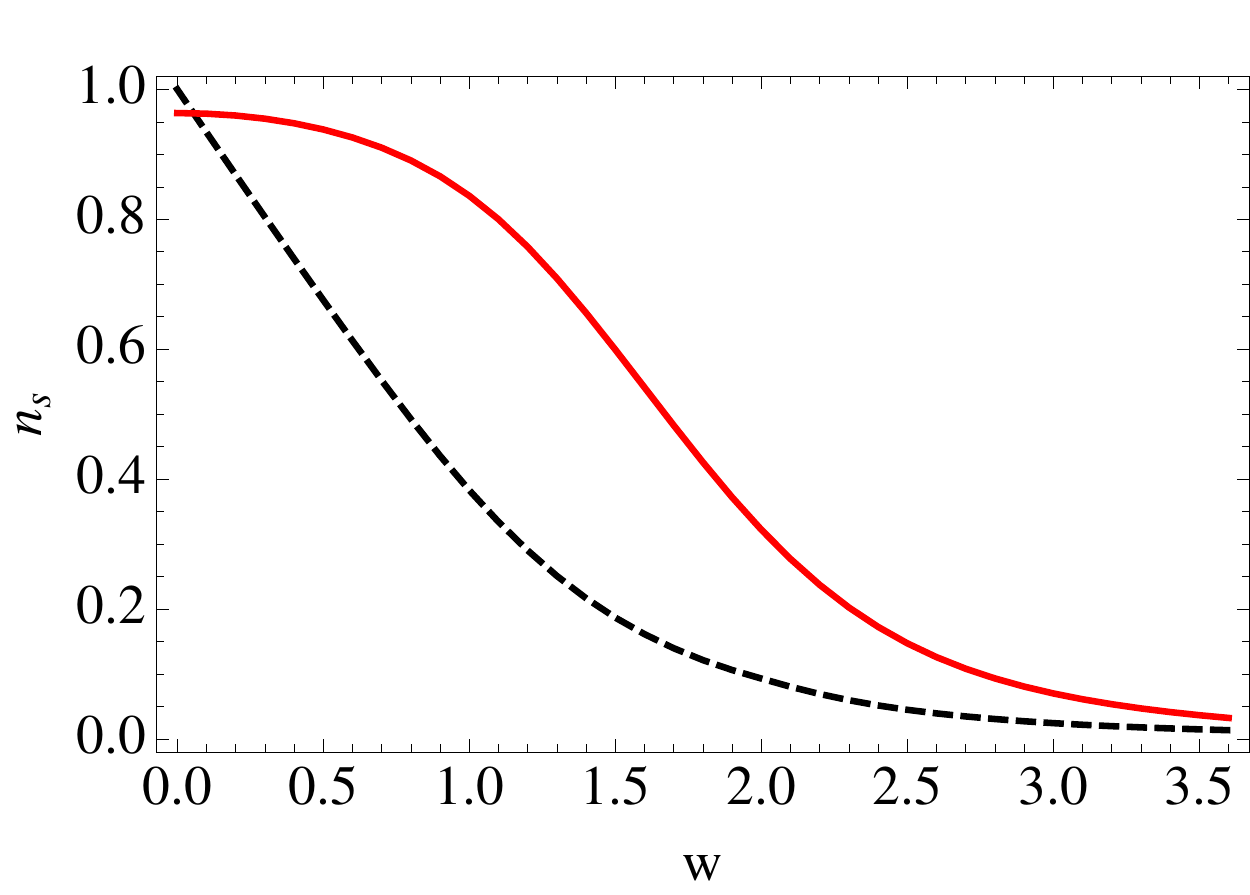}
   \caption{Superfluid density, $n_s$ (solid line)  and minimum of the condensate, ${\cal O}_{\rm min}$ 
   (dashed line), as a function of disorder, $w$, for $T/T_c^{w=0}=0.81$. The value of  ${\cal O}_{\rm min}$ 
   has been normalized to unity at $w=0$.}
   \label{fig:ns-v-disorder}
 \end{figure}

To characterize the possible appearance of islands we study the minimum of the condensate  ${\cal O}_{\rm min}$
for different 
values of the parameters that characterize the system.
In Fig.~\ref{fig:ns-v-disorder} we plot ${\cal O}_{\rm min}$ (dashed line) as function of the disorder 
strength $w$
at constant temperature $T=0.81T_c^{w=0}$.
This figure shows  that ${\cal O}_{\rm min}$ is a holographic version of the original suggestion of Ma
and Lee \cite{PhysRevB.32.5658} that strong disorder gives rise to spatial fluctuations of the order parameter
along with its suppression in comparison with its value in the clean/homogeneous system.
Moreover, for large values of the disorder strength we observe an exponential tail for ${\cal O}_{\rm min}$.
This means that we have some leaking between islands of superfluid leading to a finite value of the condensate
in between. We relate this to the fact that AdS suppresses large momentum contributions to the order parameter
as reported in \cite{Arean:2013mta,Arean:2014oaa}.


{\bf Conductivities:}
The holographic computation of conductivities requires the study of fluctuations on top of the background.
Here we focus on the conductivity of the $U(1)$  broken superconducting phase, where the perturbations couple to
the noise due to the nontrivial scalar field.
Given the symmetries of the setup, there are two different conductivities one can study:
the one in the direction parallel to the noise and that orthogonal to it. The orthogonal conductivity is
insensitive to the disorder, and we thus focus on the parallel one.

To compute the electric conductivity we consider homogeneous perturbations of
the gauge field oscillating with frequency $\omega$:
\be
A_\mu=A_\mu^{(0)}(x,z) + a_\mu(x,z)\,e^{-i\omega\,t}\,,
\label{pertdef}
\ee
and on the boundary ($z=0$) we impose $f_{ti}(x)=1$ (where $f_{\mu\nu}$ is the field strength of $a_\mu$, and 
$i$ runs over the spatial directions $x,y$), thus sourcing a constant electric field in the dual field theory. 
Then, following AdS/CFT, the conductivity is computed as
\be
\sigma_{ii}(x)=\frac{\langle J_i\rangle}{E_i}=-\frac{i \partial_z a_i(x,z)}{\omega\, a_i(x,z)} \bigg |_{z\to 0}\,,
\ee
where there is no summation over $i$, and we impose ingoing boundary conditions for $a_i(x,z)$ at the horizon.

\begin{figure}[htb]
\centering
\includegraphics[width=0.40\textwidth]{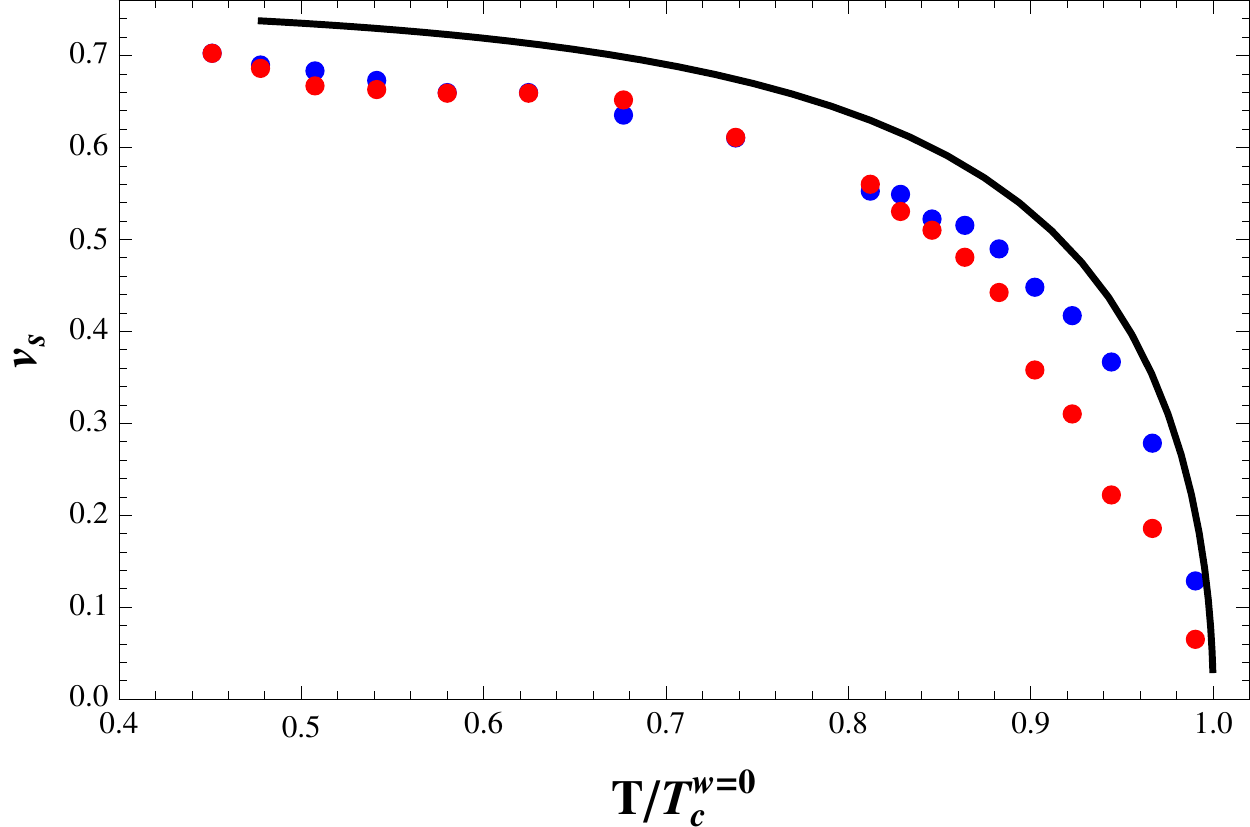}
\caption{Effective velocity $v_s(T, w)$ corresponding to the first resonance of the conductivity for $w=0.1$ 
(blue dots) and $w=1$ (red dots). The black solid line presents the result of \cite{Amado:2013xya} for the
speed of sound of the sound mode in holographic superconductors.}
\label{fig:Resonances}
\end{figure}

For the conductivity in the direction of the noise we must consider the linearized equations for the spatial
component of the gauge field $a_x(x,z)$.
Unfortunately, the $x$ dependence couples  $a_x$ to the perturbation of the temporal component of the gauge
field $a_0(x,z)$ and to the perturbations of the scalar field $\chi(x,z)+i\eta(x,z)$, and
we must then solve the four coupled linear equations of motion~\eqref{eq:fluceoms}.
Interestingly, as we will show in Fig.~\ref{fig:Resonances}, the fields $(\chi, \eta)$ show up as 
hydrodynamic  modes in the computation of conductivities.

When looking for the low frequency behavior of the conductivity we  find that,
as in the homogeneous phase, we have a pole in the imaginary part of $\sigma$.
This translates, through Kramers-Kronig rules, into a (numerically invisible) delta function
in the real part, giving
\be
\sigma\approx n_s \left(\pi\delta(\omega)+\frac{i }{\omega}\right)+\dots \,,
\ee
where $n_s$ is the superfluid density. 
Notice that for $\sigma_{xx}$, a near-boundary  analysis of the equations of motion shows that, 
due to current conservation,
the DC conductivity must be homogeneous, and therefore $n_s$ is a constant independent of $x$.

The dependence of the superfluid density on the strength of disorder is shown in Fig.~\ref{fig:ns-v-disorder},
together with the evolution of the minimum value of the condensate.
For low $w$, we verify that $n_s$ does not change much in total agreement with Anderson's theorem even if we are
describing a strongly coupled system. 
The persistence of the transport properties for weak disorder, even in strongly coupled systems, has been 
argued in the metal-insulator transition in \cite{Basko20061126}.
Fig.~\ref{fig:ns-v-disorder} shows that for large $w$, both the superfluid density and the minimum of the 
condensate decay exponentially. This illustrates the disorder-driven transition to the normal (metallic) phase.

\begin{figure}[htb]
   \centering
  \includegraphics[width=0.45\textwidth]{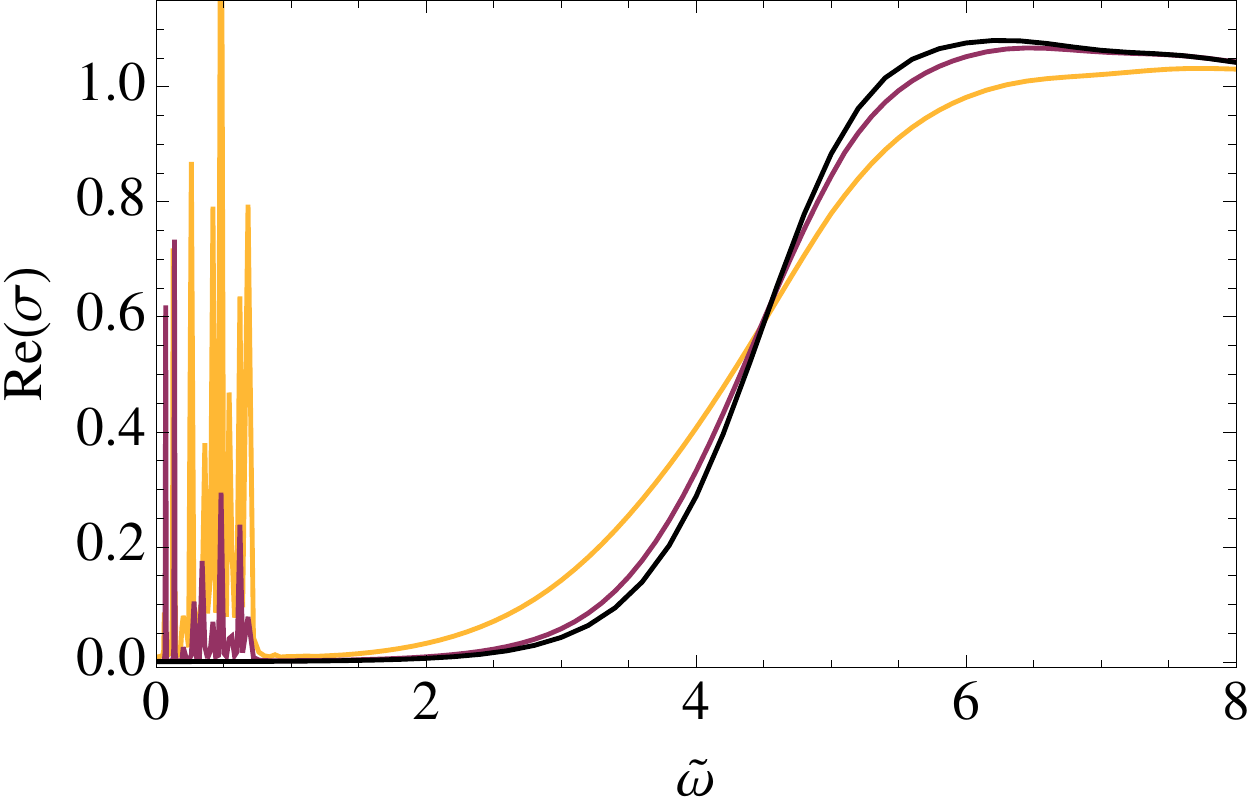}
   \caption{Real part of the AC conductivity for $T/T_c^{w=0}=0.45$. The black line corresponds to the
   homogeneous case, whereas
   the purple and yellow lines denote the disordered $w=1$ and $w=2.4$ cases respectively.
   The dimensionless frequency $\tilde\omega$ is proportional to $\omega/T$.}
   \label{fig:Conductitivies}
 \end{figure}

We present the real part of the AC conductivity in Fig.~\ref{fig:Conductitivies}, 
comparing the homogeneous case with the results for two different values of the strength of disorder.
The two key features absent in the homogeneous case are: the presence of resonances for small frequencies 
and a shift of the spectral weight at large frequencies.
The small frequency resonances are a direct result of coupling of new fluctuating fields  
for an $x$-dependent chemical potential. These resonances can be understood as related to the 
holographic quasinormal modes studied in \cite{Amado:2009ts,Amado:2013xya} and shown to affect
the conductivity at nonzero superfluid velocity in  \cite{Amado:2013aea}. 
The shift in the spectral weight clearly depends on the strength of disorder, and we view it 
as evidence of the Higgs mode associated with the spontaneous breaking of the $U(1)$ 
symmetry \cite{Sherman_Higgs_Nature}. The Higgs mode corresponds to oscillations
of the modulus of the order parameter, and it is predicted~\cite{PhysRevB.84.174522}
to cause an excess of the electrical conductivity at sub-gap frequencies. 
In clean BCS superconductors
the mass of the Higgs mode results in a gap of its contribution to the conductivity which is
of the order of the BCS gap, and thus makes it difficult to detect the Higgs mode through
its effect on the conductivity. However, it was shown in~\cite{PhysRevB.88.235108} 
that disorder suppresses that gap,
giving rise to an observable excess of the AC conductivity at sub-gap frequencies with respect
to the standard BCS prediction (based on the measurement of the energy gap of Cooper 
pairs)~\cite{Sherman_Higgs_Nature}.
The excess observed in~\cite{Sherman_Higgs_Nature} is qualitatively similar to what we observe
in Fig.~\ref{fig:Conductitivies} at frequencies $2\lesssim\tilde\omega\lesssim4$ where the conductivity
of the disordered system (yellow line) is higher than that of the clean one (black line).

To verify that the small frequency resonances in the conductivity correspond to the gapless modes studied in
\cite{Amado:2013xya}, we compute the effective velocity, $v_s$, of the first of these resonances assuming the
dispersion relation
\be
\label{eq:dispersion}
\omega=v_s (T,w)\, k\,,
\ee
and taking $k$ to be equal to $k_0$, where $k_0=2\pi/L$ is the smallest wave number in the sum (\ref{noisefunc}).
We then compare the evolution of $v_s$ with the temperature with the findings of \cite{Amado:2013xya}.
The result is plotted in Fig.~\ref{fig:Resonances} which shows that  $v_s(T,w)$ follows the standard 
temperature dependence discussed in \cite{Amado:2009ts,Amado:2013xya} (corresponding to the solid black
line in the graphic). What is new in our case is its dependence on the disorder. For small temperatures the
role of disorder is suppressed. For higher temperatures we see that $v_s$ decreases with increasing disorder
strength except very near the critical temperature, Fig.~\ref{fig:Resonances}.

{\bf Smeared phase transition:}
Finally, let us turn to the analysis of one of the most universal properties in phase transitions: 
the behavior of the order parameter close to the  transition \cite{Wilson:1973jj}. 
As a first approach to disordered phenomena, the fate of a particular clean critical point under
the influence of impurities is controlled by the Harris criterion \cite{harris1974effect}
that generalizes the standard power-counting criterion to random couplings.
As explained in
\cite{Garcia-Garcia:2015crx} our chemical potential disordered along one dimension, 
and thus perfectly correlated along the remaining 
spatial direction,
introduces relevant disorder. This places our setup in the class of systems where quenched disorder can
lead to exotic critical points where the conventional power-law scaling does not 
hold~\cite{mccoy1969incompleteness,fisher1992random,vojta2003disorder}.
Moreover, in some cases disorder has been shown to cause the formation of 
rare regions undergoing a phase transition independently from the rest of the system.
In terms of the average of the order parameter this results in a smeared phase transition
characterized by an exponential scaling~\cite{vojta2003smearing,vojta2005quantum,0305-4470-39-22-R01}.
One would expect the islands of conductivity that form in our system to play the role of
these rare regions, and consequently to smear out the phase 
transition.

\begin{figure}[htb]
   \centering
\includegraphics[width=0.40\textwidth]{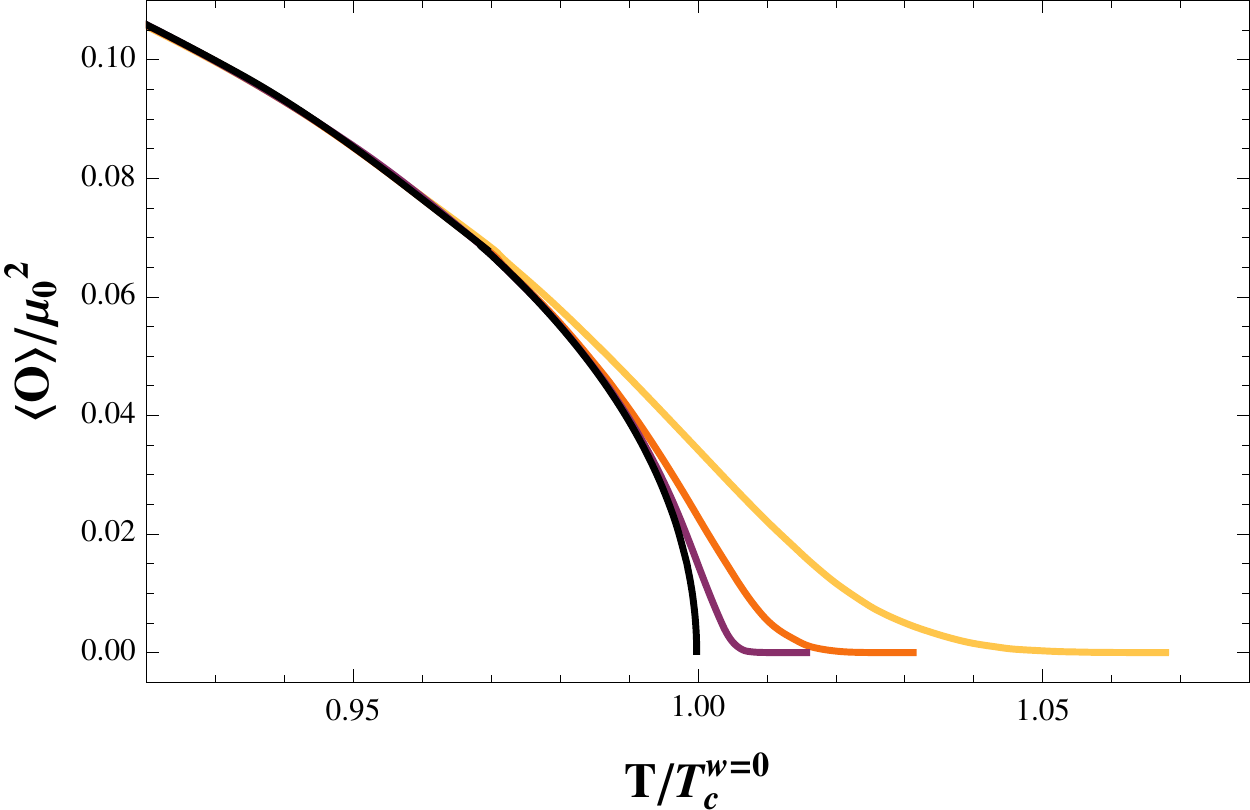}
\caption{The disorder-driven transition is smeared.
Plot of the spatial average of the condensate close to the critical temperature for $w=0$, 0.1, 0,2 and 0.4
(black, purple, orange, and yellow lines, respectively).}
\label{fig:Smeared}
\end{figure}

\begin{figure}[htb]
   \centering
  \includegraphics[width=0.40\textwidth]{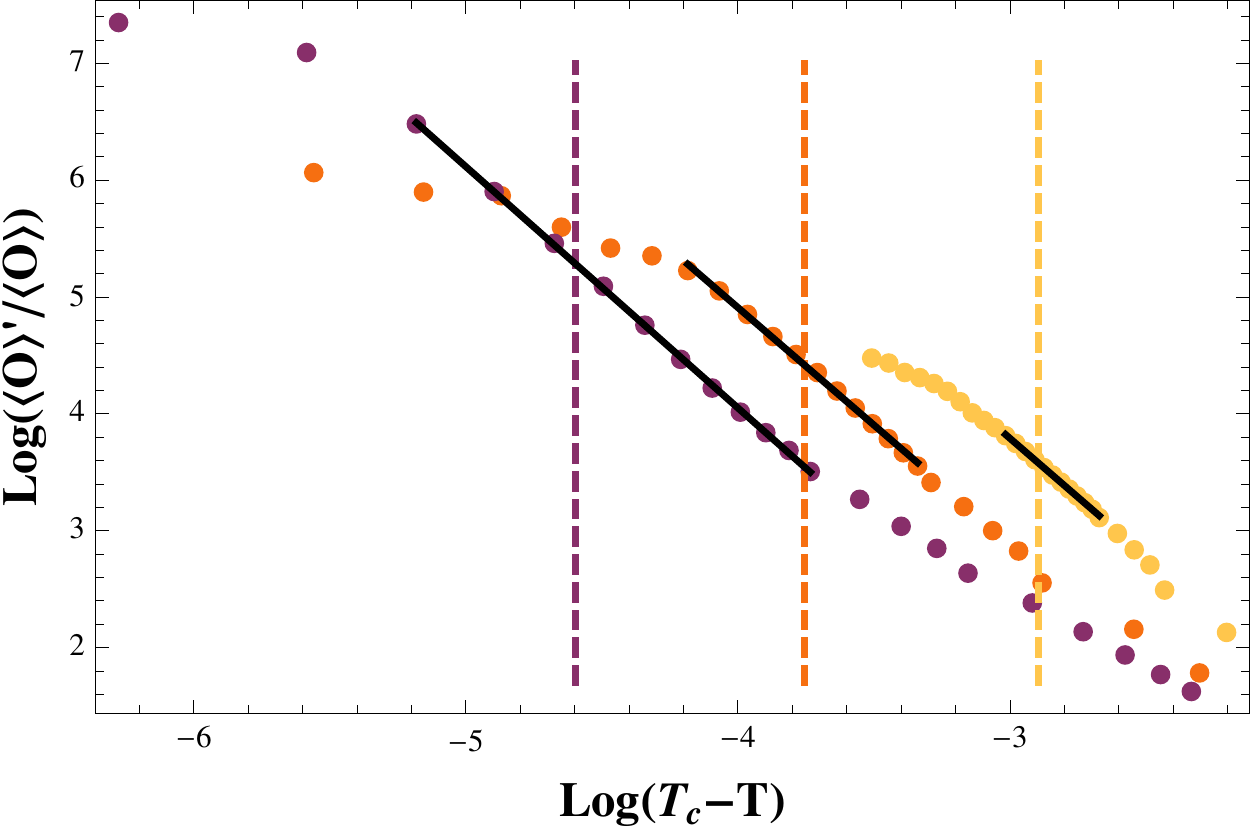}
   \caption{The disorder-driven transition is smeared. 
We plot $\log\left({1\over {\cal O}}\,{d{\cal O}\over dT}\right)$ {\it versus} $\log(T_c - T)$ for
$w=0.1, 0.2, 0.4$ (purple, orange, and yellow respectively).
The solid black lines denote linear fits to the data, and the vertical dashed lines
indicate the critical temperature of the homogeneous case $T_c^{w=0}$.}
\label{fig:fits}
\end{figure}

Fig.~\ref{fig:Smeared} shows the transition for the clean, homogeneous case \cite{Hartnoll:2008kx}, 
and for three disordered cases $(w=0.1, 0.2, 0.4)$. We confirm that for the disordered transition 
the order parameter behaves as  $\langle {\cal O}\rangle \sim \exp\left(- A|T-T_c|^{-\nu}\right)$ 
in Fig.~\ref{fig:fits}. 
Notice that $T_c$ is different for each value of $w$, and is always higher than $T_c^{w=0}$.
By taking the logarithmic derivative $\langle \mathcal{O}'(T)\rangle/\langle \mathcal{O}(T)\rangle$, 
this figure shows linear fits that allow us to determine the exponent $\nu$, 
obtaining $\nu=1.03 \pm 0.02$ independent of the disorder $w$. Explicitly, the fits in Fig.~\ref{fig:fits} 
are  $y = -2.26-2.02 x$ for  $w=0.4$,  $y= -3.13-2.01 x $ for $ w=0.2$, and $y= -4.21-2.06 x$  for $ w=0.1$, 
and notice that the 
slope of these fits corresponds to $-(\nu+1)$.
The fits also show that the coefficient $A$ of the expression above
increases with the disorder strength $w$,
in agreement with the optimal fluctuation  theory predictions~\cite{vojta2003smearing,0305-4470-39-22-R01}.
In particular, after assuming $\nu \approx 1$, 
these fits result in the following values of $A$: 0.01 at $w=0.1$, 0.04 at $w=0.2$,
and 0.1 at $w=0.4$.
The ranges of the fits, shown by solid black lines in Fig.~\ref{fig:fits}, are determined on one side by the 
region where the condensate behaves as in the homogeneous case  which  as  Fig.~\ref{fig:Smeared} shows is for 
temperatures around $.95T_c$. On the other side the bound arises from the numerical instability introduced by
exponentially small values of the condensate. 
Notice that the points at temperatures above $T_c^{w=0}$ are rare, and many realizations are needed in order 
to have enough statistics to capture their behavior.
Moreover, those points correspond to very low values of the condensate, and small 
errors in $\langle \mathcal{O}(T)\rangle$ affect strongly the value of  
$\langle \mathcal{O}'(T)\rangle/\langle \mathcal{O}(T)\rangle$. 
Finally, one should notice that the UV cutoff $k_*$ in our disordered chemical potential 
bounds the disorder distribution. Consequently, one expects to find, as we do, a finite disordered critical 
temperature $T_c^w$ above which only the normal phase 
exists~\cite{vojta2003smearing,0305-4470-39-22-R01}.

{\bf Conclusions}\\
{\it Superconducting islands in holography:} We argued that the formation of islands in the context of holographic
disorder-driven superconductors is, as in experimental and numerical analyses, the central mechanism at play in 
the transition. 
This behavior is responsible for the enhancement of superconductivity reported in~\cite{Arean:2013mta,Arean:2014oaa}.
The fact that the islands are coupled via the Josephson effect, as suggested 
originally in \cite{PhysRevLett.87.087001},
causes the exponential decay in the tail of the superfluid density as shown in Fig.~\ref{fig:ns-v-disorder}.

{\it Disorder in strongly interacting systems:}  The interplay between disorder and interactions has been the 
subject of vigorous studies in the condensed matter literature initiated most recently by the work of
Basko {\it et al.}~\cite{Basko20061126}. In the series of works generated by \cite{Basko20061126}, 
(see for example \cite{PhysRevB.75.155111,huse2014phenomenology, nandkishore2014many}) it is shown that for 
sufficiently strong disorder, weak interactions do not change the nature of the localized 
phase~\cite{chandran2015constructing,ros2015integrals}. The present work is in the complementary region, 
where interactions are strong to begin with, and disorder is increased.  
We have established that weak disorder does not destroy the holographic superconducting state, in particular,
Fig.~\ref{fig:ns-v-disorder} shows that the superfluid density is largely unaffected for small disorder.
Instead, for strong disorder the formation of islands suppresses the superfluid density driving the system to
the normal phase.

{\it Optical conductivity:}  We have studied the AC conductivity for disordered holographic superconductors, 
identifying the new low-frequency resonances with quasinormal modes of the holographic superconductor. 
Moreover, the conductivity
displays a disorder-dependent shift of the spectral weight highly suggestive of a massive Higgs excitation.
Strong experimental evidence in favor of a  Higgs mode in strongly disordered superconductors close to the 
quantum phase transition has been recently reported in \cite{Sherman_Higgs_Nature}.

{\it The disorder-driven transition is smeared:} We have shown that under the influence of disorder the 
superconductor-metal transition changes from power law, mean field to a smeared transition of the 
form $\exp(- A\, |T-T_c|^{-\nu})$, with $\nu =1.03\pm 0.02 $. We hope that our results motivate other
approaches to a quantitative description of this transition.

{\it Future directions :} Understanding the properties of conductivities in the language of modes helps 
us compare with more traditional condensed matter methods directly.  Building a holographic disordered
superconducting thin film will help us tackle the  Higgs mode in the strongly disordered superconductor 
close to the quantum phase transition \cite{Sherman_Higgs_Nature}. Ultimately, due to the universality of 
critical exponents \cite{Wilson:1973jj}, we hope that our holographic approach provides a direct path to  
disorder-driven critical exponents.

{\bf Acknowledgments}
It is a great pleasure to  thank Nico Nessi for useful discussions.
We thank the Galileo Galilei Institute for Theoretical Physics for hospitality and the INFN (Firenze)
for partial support. 
D. A. is supported by GIF, grant 1156.
We thank Zampolli for being a shelter of flavor.
D.A.  and I.S. thank the ICTP for hospitality at various stages of this collaboration.
D.A. thanks the FROGS for unconditional support.

\appendix

\section{Supporting material}

\begin{figure}[htb]
   \centering
   \includegraphics[width=0.40\textwidth]{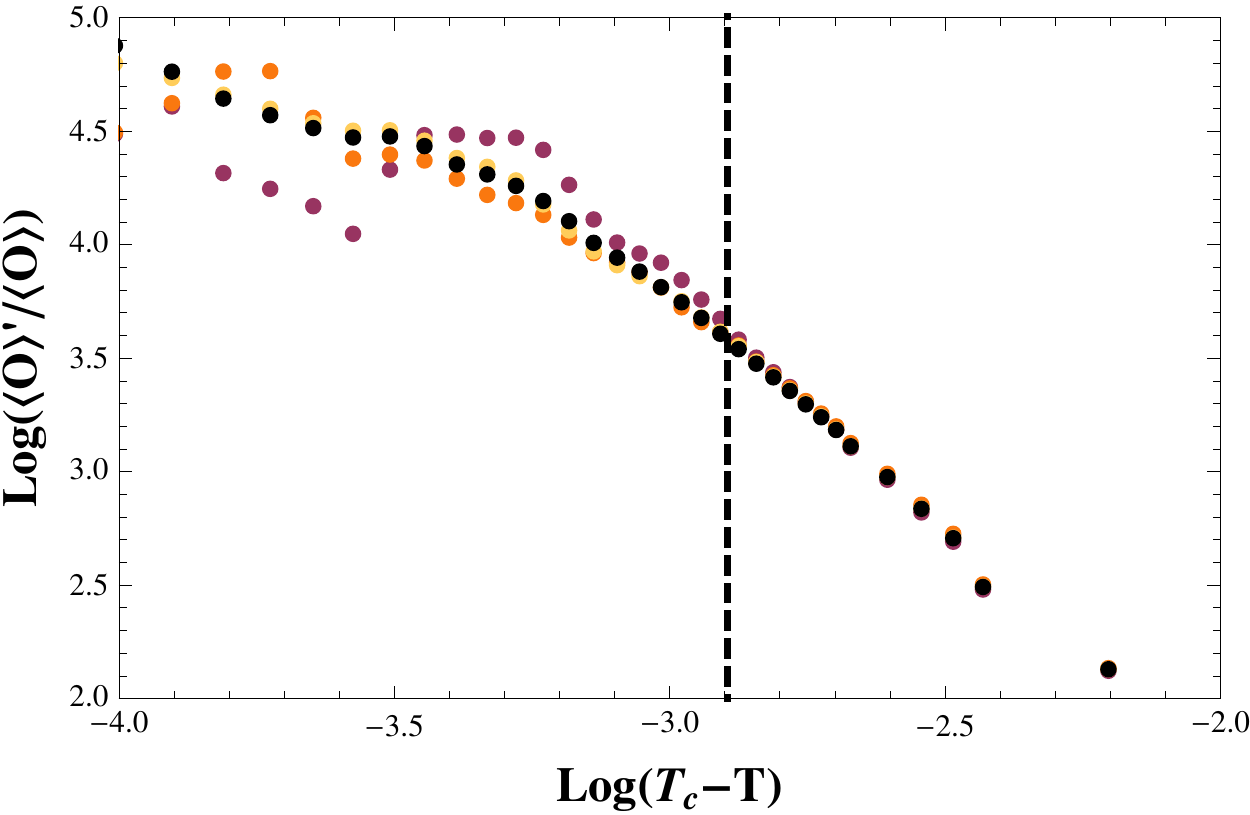}
   \caption{The condensates for $w=0.4$ with four  different numbers of realizations: purple 
   corresponds to 10 realizations, orange to 25, yellow to 75, and black to 150.  
Notice that except for the leftmost, high temperature, data points, the value of the condensate stabilizes 
for a number of realizations above 25.}   
\label{fig:realizations}
\end{figure}

{\bf Numerical stability:}
Given the aleatory nature of our calculation, in this appendix we show the stability of the result 
of the value of $\nu$ against the number of realizations. Namely, Fig.~\ref{fig:realizations} shows 
that the result stabilizes as we increase the number of realizations of disorder, {\it i.e.} the number
of times we generate a series of random phases, $\delta_k$ in Eq. (\ref{noisefunc}). 
Most of our simulations for determining $\nu$ involve over 100 realizations.

\begin{figure}[htb]
\centering
\includegraphics[width=0.40\textwidth]{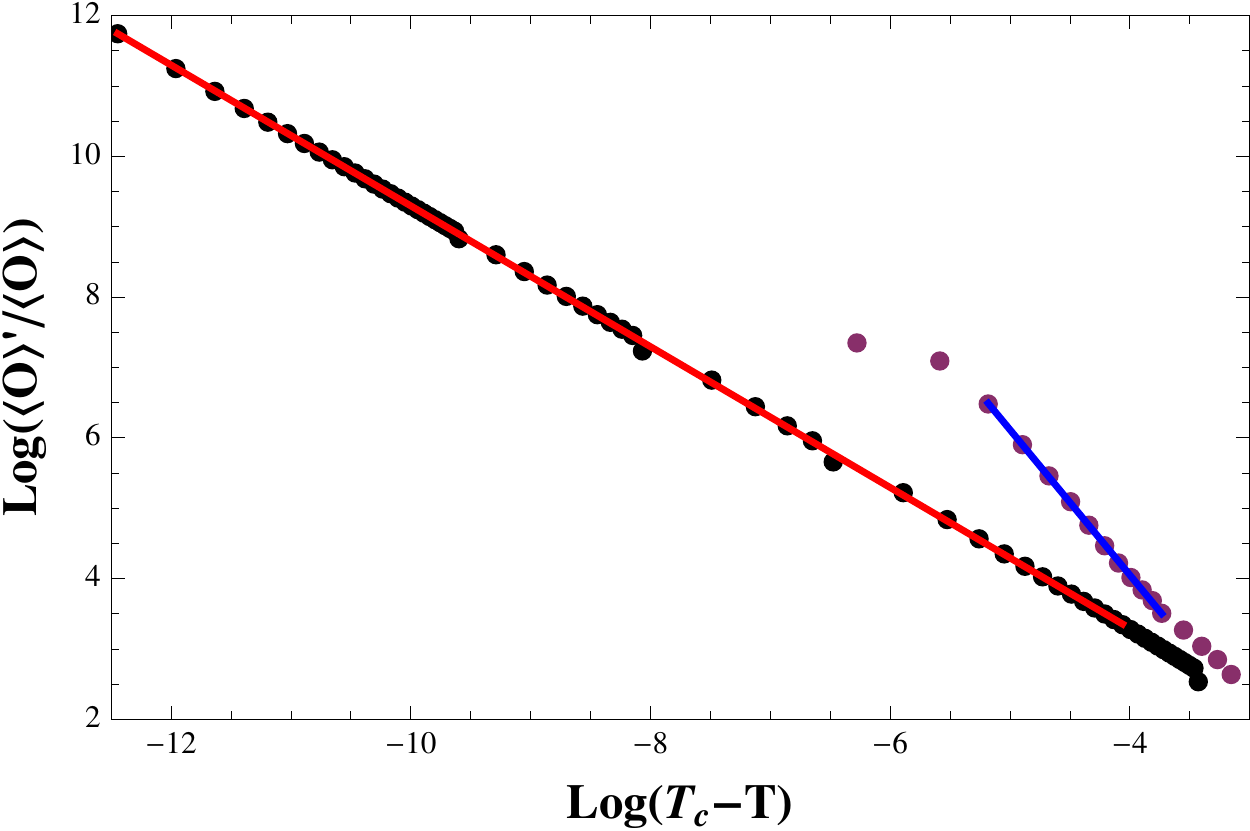}
\caption{The homogeneous condensate (black dots) and the smeared condensate for $w=0.1$ (purple). The red,
and blue lines correspond to the linear fits $y=-0.71-1.00x$, and $y= -4.21-2.06 x$ respectively.}
   \label{fig:HomoandSmeared}
 \end{figure}

{\bf Homogeneous versus smeared condensate:}
In Fig.~\ref{fig:HomoandSmeared} we show that the homogeneous transition follows a power law.
We present it in the same plot as a smeared transition to 
illustrate how our numerical analysis distinguishes between the homogeneous and the disordered
transition. 
Note that for an homogeneous condensate ${\cal O}_h = A_h|T-T_c|^\beta$, with
$\beta=1/2$ \cite{Hartnoll:2008kx}, hence one can write
\be
\log \left(\frac{{\cal O}'}{\cal O}\right)=- \log |T-T_c| + {\rm cons}\,,
\ee
while for the disordered case ${\cal O} = A_s \exp(- A |T-T_c|^{-\nu})$, and one has
\be
\log \left(\frac{{\cal O}'}{\cal O}\right)=- (\nu +1)\log |T-T_c| + {\rm cons}\,.
\ee

{\bf Equations of the perturbations:}
The equations of motion for the perturbations relevant for computing the electric conductivity read:
\begin{widetext}
{\small
\begin{subequations}
\begin{align}
& 2\left(i\,\omega\, \eta+a_0\,\psi+2\,\phi\,\chi\right)\psi 
- z^2\left(f\,\partial_{zz}a_0+i\,\omega\,\partial_x a_x+\partial_{xx}a_0\right)=0\, ,\\
&\frac{z^2}{f}\left( \omega^2 \,a_x+f\,f'\,\partial_z a_x
+f^2\,\partial_{zz}a_x-i\,\omega\,\partial_{x}a_0\right)
-2\left(\psi^2 a_x-\psi\,\partial_x\eta+\partial_x\psi\,\eta\right)=0\, ,\\
&-2\left(i\,\omega\,\eta+\psi\, a_0\right)\phi-z^2\phi^2\,\chi
+\left(-z^2\,\omega^2+m^2\,f\right)\chi-z\,f\left(-2\,f\,\partial_z\chi+z\,f'\,\partial_z\chi
+z\,f\,\partial_{zz}\chi+z\,\partial_{xx}\chi\right)=0\, ,\\
&\left[z^2\left(\omega^2+\phi^2 \right)-m^2\,f \right] \eta   
-2\,x\,f^2\, \partial_z\eta+z^2\left[-i\,\omega\,a_0 \,\psi-2\,i\,\omega\phi\,\chi
+f\left( f'\,\partial_z\eta+f\,\partial_{zz}\eta
-\psi\,\partial_x a_x-2\,a_x\,\partial_x\psi+\partial_{xx}\eta   \right) \right] =0\, ,
\end{align}
\label{eq:fluceoms}
\end{subequations}}
\end{widetext}
subject to the constraint:
\be
-i\,z^2\,\omega\,\partial_z a_0+f\left(2\,\psi\,\partial_z\eta-2\,\eta\,\partial_z\psi-z^2\,\partial_{xz}a_x\right)=0\, .
\ee

\bibliography{CMbib}

\end{document}